\documentclass[11pt, a4paper]{article}
\pdfoutput=1 

\usepackage{jheppub} 

 \DeclareUnicodeCharacter{2212}{-}
\usepackage{float,comment}
\usepackage{slashed}
\usepackage{xcolor}
\usepackage{mathrsfs}
\usepackage{caption}
\usepackage{subcaption}
\usepackage{lipsum} 
\usepackage[english]{babel}
\usepackage{lineno, blindtext}

\newcommand{\VLL}{L}
\newcommand{\tVLL}{\tilde{L}}
\newcommand{\whizard}{{WHIZARD}}

\preprint{
\begin{minipage}{5cm}
\small
\flushright
CTPU-PTC-24-33
\end{minipage}
}

\title{Investigating vector-like leptons decaying into an electron and missing transverse energy in e$^{+}$ e$^{-}$ collisions with $\sqrt{s} = 500$ GeV at the ILC}

\author[a,b]{Y. Mahmoud,\footnote{yehia.abdelaziz@bue.edu.eg}}
\author[c]{J. Kawamura,\footnote{junkmura13@gmail.com}}
\author[b]{H. Abdallah,}
\author[b]{M. T. Hussein,}
\author[a]{S. Elgammal}

\affiliation[a]{Centre for Theoretical Physics, The British University in Egypt, P.O. Box 43, El Sherouk City, Cairo 11837, Egypt.}
\affiliation[b]{Physics Department, Faculty of Science, Cairo University.}
\affiliation[c]{
Center for Theoretical Physics of the Universe, Institute for Basic Science (IBS), Daejeon
34051, Korea}

\abstract{

This analysis focuses on probing the lepton portal dark matter 
using Monte Carlo simulated samples from electron-positron collisions 
at the International Linear Collider (ILC) 
of 500 GeV center of mass energy 
with an integrated luminosity of 1000 fb$^{-1}$. 
The study examines a benchmark scenario 
where the dark matter is a scalar particle produced as a daughter particle 
of the vector-like lepton. 
The signal topology consists of missing transverse energy and dilepton. 
If no new physics is discovered,
the study sets 95\% confidence level exclusion limits 
on the mass of vector-like leptons.
}


\begin{document}
\maketitle
\flushbottom

\newcommand{\RNum}[1]{\uppercase\expandafter{\romannumeral #1\relax}}


\section{Introduction}
\label{sec:intro}

Dark Matter, a non-visible matter that pervades our universe, 
is one of the most important phenomena in physics. 
The recent astrophysical observations \cite{planck2015,planck2018} suggest that it constitutes about 25\% of the universe's energy, 
while visible matter constitutes only 5\%. 
Evidences for the existence of dark matter have been found through various indirect detection experiments, such as the galaxy rotation curves, gravitational lensing 
and bullet cluster experiments~\cite{bullet_cluster,dm_evidence}.  
Since these evidences are all through the gravitational effects, 
we do not know the nature of the dark matter as a particle.  
A promising way to probe the dark matter 
is through the direct detection of dark matter particles, 
which are assumed to be produced from collisions of high-energy particles 
in particle accelerators \cite{CMS_dark_matter,ATLAS_dark_matter}.

While the Standard Model (SM) of particle physics is considered 
as the most successful theory of particle physics, 
it fails to account for the dark matter. 
Therefore, there is a need for models beyond the standard models (BSMs) 
that can describe dark matter particles and their interactions. 
A suitable dark matter candidate should be able to achieve the observed relic density 
of dark matter in the universe. 
This requirement is fulfilled by weakly interacting massive particles (WIMPs), 
which have successfully explained the correct observed relic density 
for dark matter in the universe,  known as the WIMP miracle~\cite{wimps}.

The lepton portal dark matter model proposed in \cite{Chang:2014tea,Bai:2014osa} 
involves a singlet dark matter under the SM gauge group.
The dark matter particles annihilate into the SM leptons through 
the $T$-channel exchange of the vector-like leptons whose
left and right-handed components transform in the same way 
under the electroweak (EW) gauge group of the SM. 
Vector-Like leptons are ubiquitous in BSM models, such as 
supersymmetric models \cite{susy1,susy2,susy3,susy4}, 
grand unification theories \cite{gut1,gut2,gut3}, 
extra dimensional models \cite{extra_dim1,extra_dim2}. 
and models for the dark matter \cite{vll_dm1,vll_dm2,vll_dm3,vll_dm4}. 
They can also explain the mass hierarchy between different generations 
of particles \cite{hierarchy1,hierarchy2,hierarchy3}, 
and the discrepancy between the measured 
and the predicted anomalous magnetic moment of the muon \cite{susy3,muon_moment1,muon_moment2,muon_moment3,muon_moment4, lpdm, Kawamura:2022uft}. 
Searches for singlet and doublet vector-like tau leptons were performed by the CMS in leptonic and hadronic final states \cite{cms_vll1,cms_vll2,cms_vll3}, where the doublet vector-like tau lepton mass was constrained in the mass range 120-790 GeV.
Vector-like Leptons were also studied by the ATLAS experiment in events with hadronic final states in \cite{vll_atlas}. The leptonic channels at the ATLAS with the Run-2 data are used to recast the limit on the vector-like leptons decaying to the SM leptons in the second generation in \cite{Kawamura:2023zuo}. The latest study on singlet Vector-Like leptons constrained the Yukawa coupling to be as low as [0.032, 0.098] for masses of Vector-Like leptons between 101 and 463 GeV.

In this study, we will focus on the vector-like lepton which decays to a final state of SM electrons and missing energy. 
The study is performed for different scenarios of the mass splitting between the vector-like lepton and the scalar dark matter. 
The details of the production of the vector-like lepton are explained in detail in section \ref{section:model}. 
Since the dark matter particle does not interact with the materials in the detector, it contributes to a missing transverse energy $E_{T}^{miss}$ which is calculated from the vector sum of the transverse energy of visible particles.  
By doing so,  we can identify the lepton portal dark matter through the signature of two leptons and missing transverse energy.  

We suggest a study focused on the pair production of vector-like leptons from electron-positron collisions at the International Linear Collider (ILC) to search for dark matter.
The ILC, an electron-positron collider, is one of the proposed future particle colliders. This collider is proposed in Japan and is expected to operate at varying energies of 250 GeV, 350 GeV, 500 GeV, and 1 TeV. While other future lepton colliders such as FCC-ee \cite{FCC} and CLIC \cite{CLIC} also exist, the focus of our study will be on the ILC.
One of the significant differences between lepton and hadron colliders is that lepton colliders can provide longitudinally polarized beams. The ILC, specifically, is anticipated to have 80\% polarization for the electron beams and 30\% polarization for the positron beams, as per the sources cited in \cite{ilc,ilc2,ilc3,ilc4}.

The rest of the paper is organized as follows. 
We briefly introduce the model in Sec.~\ref{section:model}, 
and then our method for the Monte-Carlo (MC) simulation 
is explained in Sec.~\ref{section:MCmethod}. 
The selection criteria of the generated events  
are shown in Sec.~\ref{section:MCandDat}. 
The results are shown in Sec.~\ref{section:results}. 
Section~\ref{section:summary} is devoted to summary.

%
\section{The simplified model}
\label{section:model}

\begin{figure}
\begin{subfigure}[b]{0.45\textwidth}
        \centering
    \includegraphics[width=7cm]{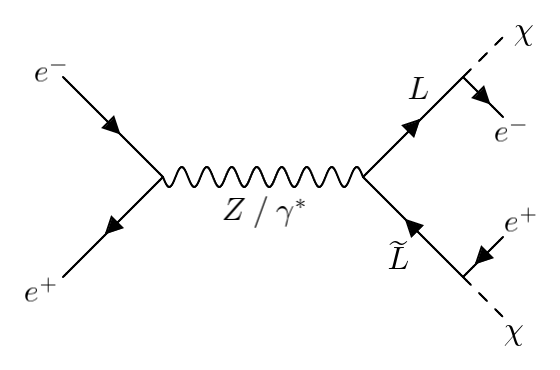}
    \qquad
   \caption{}
   \label{figure:schannel}
\end{subfigure}
\begin{subfigure}[b]{0.6\textwidth}
        \centering
    \includegraphics[width=5cm]{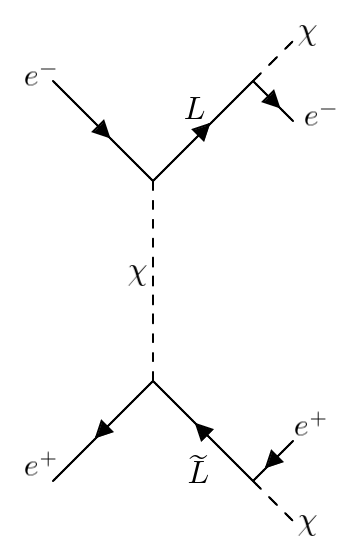}
    \qquad
    \caption{}
    \label{figure:tchannel}
\end{subfigure}
   \caption{
   Feynman diagrams of the vector-like lepton pair production 
   for (a) S-channel and (b) T-channel processes, 
   followed by the decay to the dark matter.}
    \label{figure:fig1}
\end{figure}

We shall study the lepton portal dark matter model in which
the dark matter particle is singlet under the standard model gauge group. 
The dark matter can be scalar or fermion. 
For the case of this study, we shall work on the scalar dark matter, 
and hence the mediator is the vector-like leptons. 
The full details of the model are presented in \cite{lpdm}.
Since $\chi$ is SM singlet, 
it will couple to the SM through Yukawa couplings ($\lambda^i_L$) 
through new extra leptons whose gauge quantum numbers 
are $(2, −1/2)$, 
under the EW gauge group  $SU(2)\times U(1)$. 
To avoid anomalies, the extra leptons are made vector-like. 
We denote the EW doublet 
vector-like leptons as $L$. 
Also, an additional $Z_2$ or global $U(1)$ symmetries are imposed to stabilize the dark matter. All the SM model fields trivially transform under this additional symmetry while the dark matter and the extra leptons transform non-trivially.

The Lagrangian of the vector-like masses and Yukawa interactions are given by 
\begin{align}
  -\mathcal{L} 
  = M_L \bar{L}_L L_R 
          +\lambda_L^i X\bar{\ell}_{L_i} L_R 
          .
\end{align}
The vector-like leptons are assumed to exclusively couple to 
either the first or second generation of the SM leptons to avoid lepton flavor violations strongly constrained by the experiments. 
We focus on the doublet vector-like lepton 
which couples to the electron. 
In this case, there are two parameters relevant 
to the following analysis, 
namely the vector-like lepton mass $M_L$ 
and the portal Yukawa coupling $\lambda_L:=\lambda_L^1$. 
Here, the mass $M_L$ is understood as the mass of the charged component 
in the doublet which is only relevant for the dilepton plus $E_T^{miss}$ signal studied in this work. 
The light doublet vector-like lepton
can increase the mass of the $W$ boson~\cite{Kawamura:2022uft,Kawamura:2022fhm} 
which is measured to be larger than the SM prediction 
by the CDF~\cite{CDF:2022hxs}.

For our case, we will study the production of the vector-like lepton $L$ pairs 
and their consequent decay to a SM charged electron and a scalar dark matter. 
The vector-like leptons 
can be produced from the S-channel process in the Feynman diagram 
in figure \ref{figure:schannel} mediated by $Z/\gamma$, 
or a T-channel process in the Feynman diagram in figure \ref{figure:tchannel} 
mediated by the scalar dark matter particle $\chi$. 
The contribution from the T-channel process depends on the strength of the coupling constant $\lambda_L$ and is negligible for small values of $\lambda_L$.
The typical signature for this process is two electrons plus large missing transverse energy from the stable dark matter particles.
The free parameters in the model are the dark matter mass $(M_{\chi})$, the vector-like lepton mass $(M_{\VLL})$, and Yukawa coupling $(\lambda_{L})$.

\begin{figure}[t]
\centering
\selectfont
\resizebox*{10cm}{!}{\includegraphics{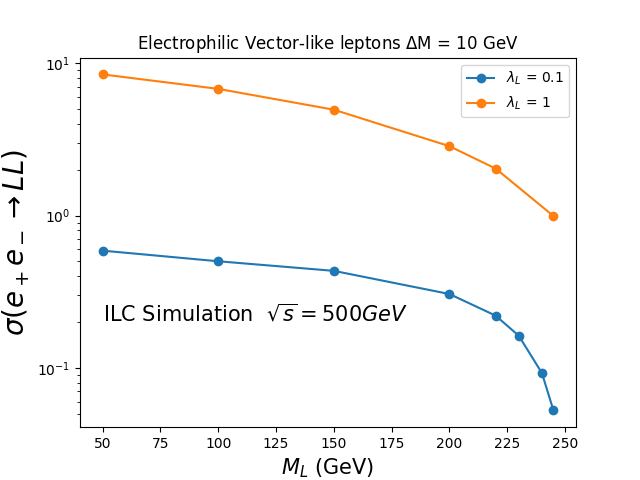}}
  \caption{Values of cross-section in pb at $\sqrt{s} = 500$ GeV at the ILC.}
  \label{figure:fig2}
\end{figure}

\section{MC simulation of signal and background}
\label{section:MCmethod}

\subsection{Monte Carlo simulation of the signal}
\label{section: MC simulation of signal and background}

To evaluate how the vector-like leptons signal would look in the ILC, 
we generated signal events using MC event generator WHIZARD-3.1.4 \cite{whizard}, which is a multipurpose event generator that has its code but uses interfaces with other packages to complete different tasks. 
O’MEGA \cite{Omega} is used for the tree-level matrix element calculation and the MC integration package VAMP is used for phase space integration.

The process of event generation was done for electron-positron beams at energies of 250 GeV each.  
For linear lepton colliders, electrons lose energy while they are being accelerated radiating photons in the process. 
The initial state radiation (ISR) from leptons has an intrinsic implementation in {\whizard} that includes all orders of soft and soft-collinear photons as well as up to third order in hard-collinear photons.
The second source of photons in the initial state is due to the macroscopic interaction of bunches (Beamstrahlung) which leads to distortion of the beam spectrum. 
For an exact simulation of the beam spectrum, Beamstrahlung must be taken into account. This is done using GuineaPig++ \cite{guinea_pig1,guinea_pig2,guinea_pig3} interfaced to \whizard.

Linear lepton colliders also can provide longitudinally polarized electron and positron beams. The ILC is expected to provide 80\% polarized electron beams and 30\% polarized positron beams. This is taken into account in {\whizard} by specifying the beam polarization for each of the electron and positron beams. 
{\whizard} can polarize the initial state fully or partially by assigning a non-trivial density matrix in the helicity space. The program used for showering and hadronization is PYTHIA6 \cite{pythia6}. For the simulation of the interaction of particles with the material of the detector and the readout system, a fast simulation algorithm (DELPHES) is implemented \cite{delphes}. 
For our study, we used the Delphes card for the international large detector (ILD). 

We generated signal events by producing vector-like leptons in pairs, which later decayed into dark matter and SM-charged electrons. The decay process followed the formula: 
\begin{align}
e^+ e^- \rightarrow \VLL {\tVLL},~~\VLL 
\rightarrow e^{-} \chi,~~{\tVLL} \rightarrow e^{+} \chi.
\end{align}
The study considered different mass-splitting scenarios between the vector-like leptons and the dark matter $\Delta M =5, 10, 100$ GeV. 

We investigated the production of vector-like leptons through T-channel production mediated by dark matter, 
in addition to the S-channel production.
The decay of $\VLL$ to electron and $\chi$ has a branching ratio of unity, 
but the $\chi$ mass has a considerable effect on the contribution 
from the T-channel process. 
If the Yukawa couplings are large enough, the DM mass has a significant contribution to the cross-section. We performed two sets of event generation, one with a small Yukawa coupling of $\lambda_L = 0.1$ and the other with a large Yukawa coupling equal to $\lambda_L = 1$.
For the small Yukawa coupling case, the t-channel process contribution is negligible, while for the large Yukawa coupling case, there is a significant contribution from the T-channel process. We plotted the production cross-sections for various dark matter masses and $\lambda_{L} = 1$ against the mass of the vector-like lepton in figure \ref{figure:fig2}.
We see that the the cross section in the large Yukawa coupling case 
is much larger than that in the small Yukawa coupling case, 
and thus the T-channel process is dominant in the former case. 
We also note that the production cross section 
for the vector-like lepton couples to a heavy flavor lepton, 
i.e. muon or tauon, will be almost the same 
for the case of the small Yukawa coupling 
in which the T-channel process is negligible at $e^+e^-$ colliders.

\begin{figure}[t]
\centering
\begin{subfigure}[b]{0.45\textwidth}
\centering
\includegraphics[width=\textwidth]{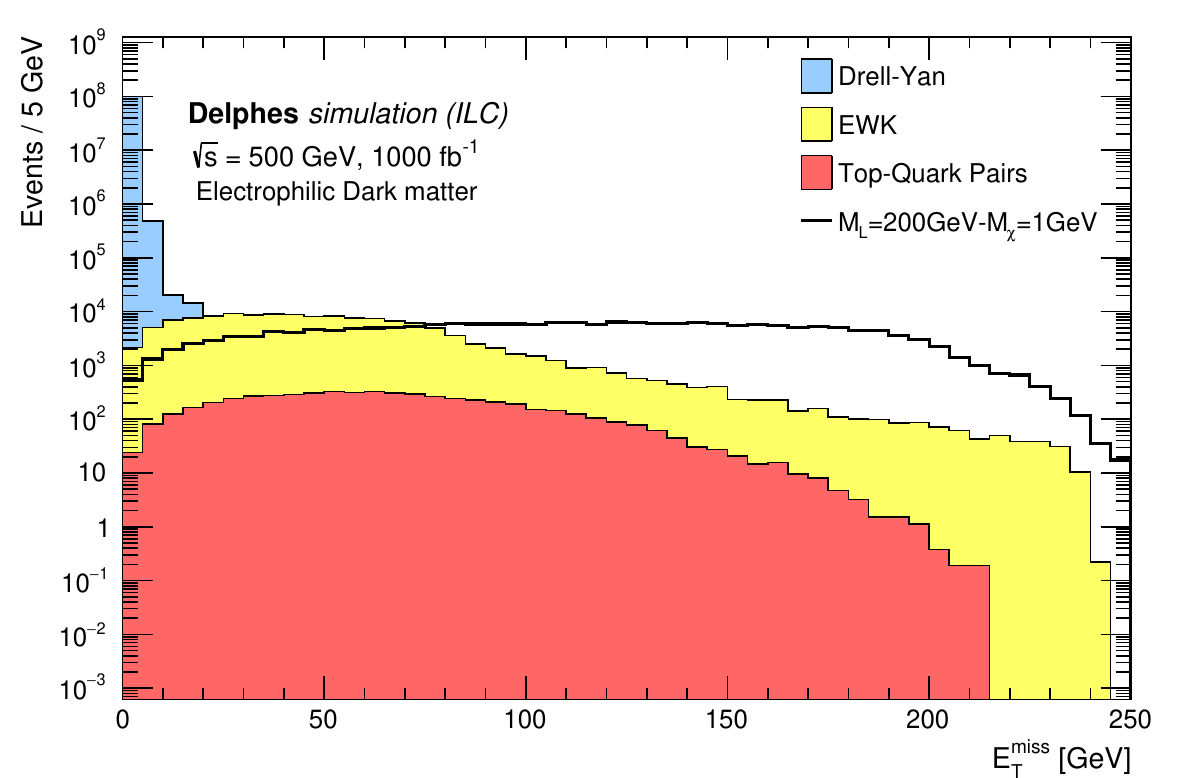}
\caption{}
\label{figure:pfmet_bef_ele}
\end{subfigure}
\begin{subfigure}[b]{0.45\textwidth}
\centering
\includegraphics[width=\textwidth]{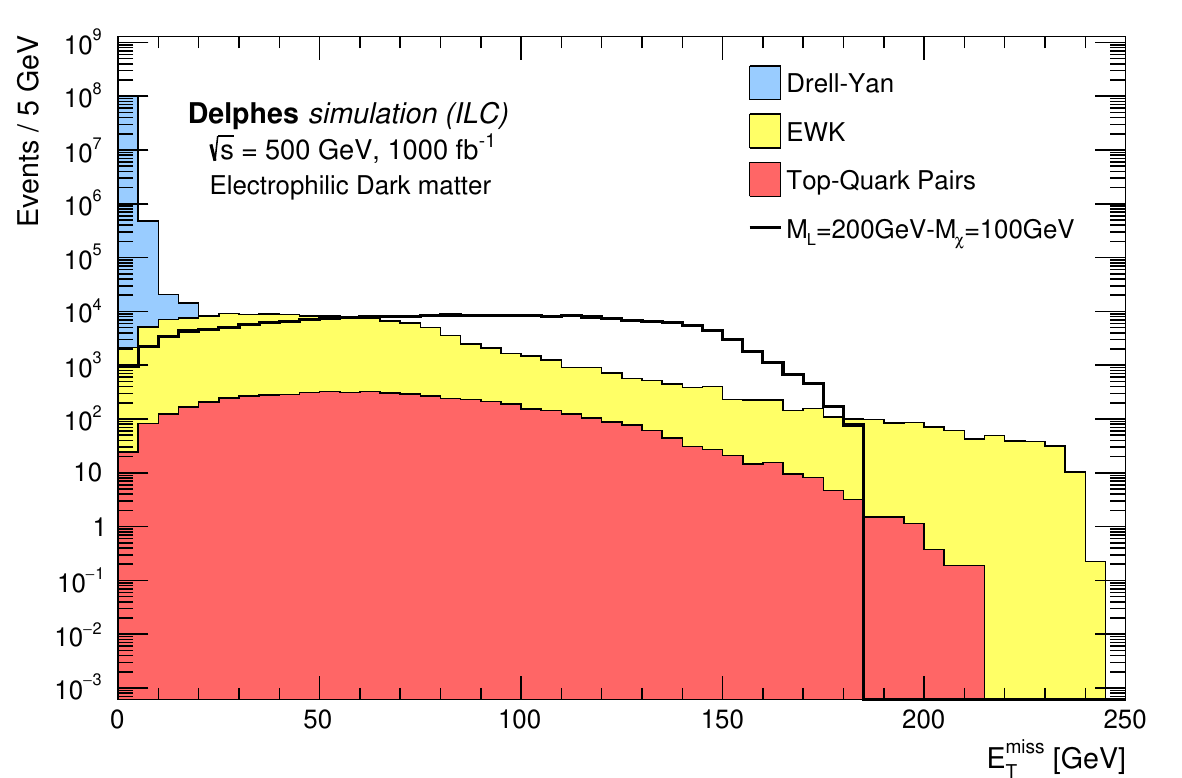}
\caption{}
\label{figure:pfmet_bef_ele}
\end{subfigure}
\begin{subfigure}[b]{0.45\textwidth}
\centering
\includegraphics[width=\textwidth]{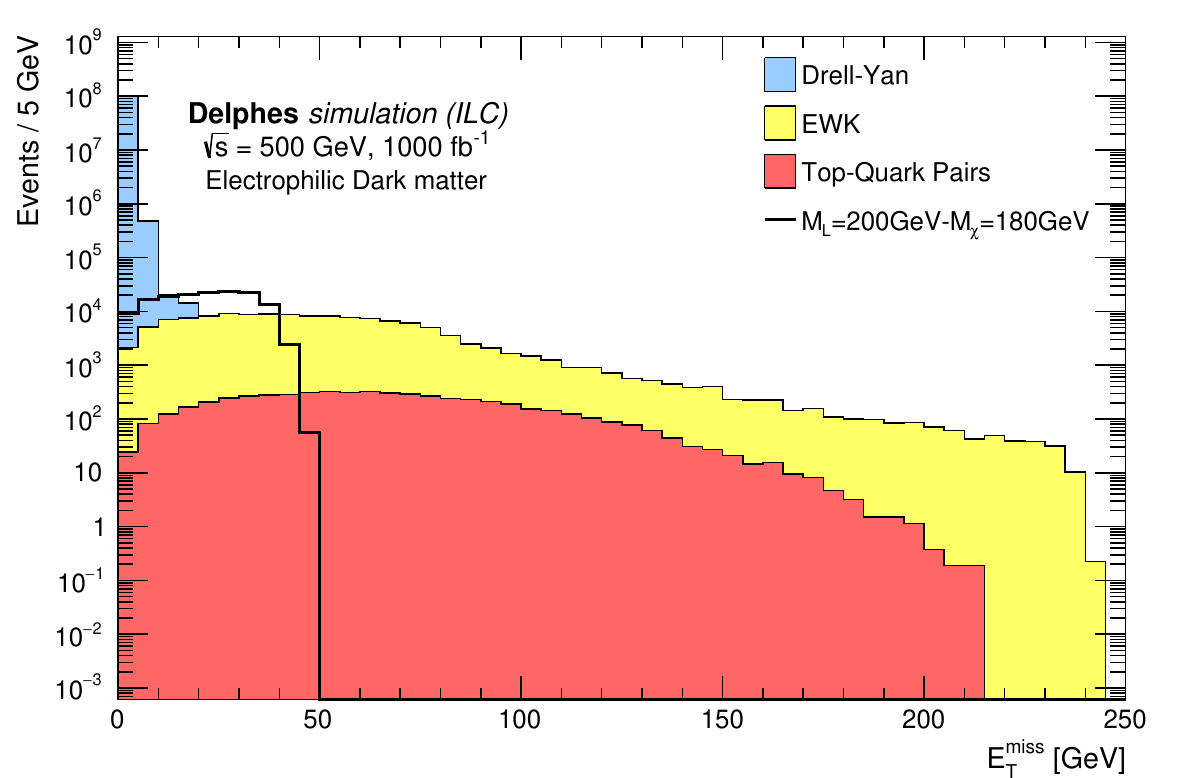}
\caption{}
\label{figure:pfmet_bef_ele}
\end{subfigure}
\begin{subfigure}[b]{0.45\textwidth}
\centering
\includegraphics[width=\textwidth]{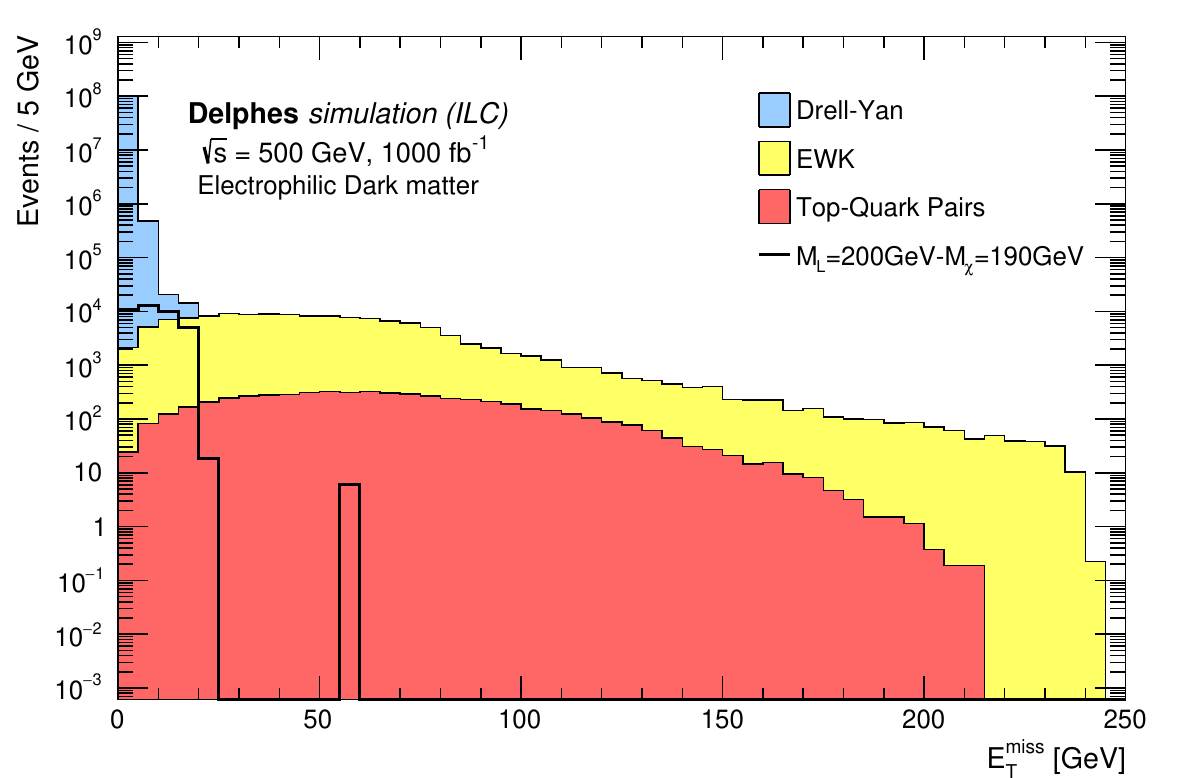}
\caption{}
\label{figure:pfmet_bef_ele}
\end{subfigure}
\caption{The missing transverse energy spectrum, for events passing the pre-selection cuts listed in table \ref{table:tab3}, of the estimated SM backgrounds together with the lepton portal dark matter 
for vector-like lepton mass of 200 GeV and scalar dark matter mass 
of (a) 1 GeV,(b) 100 GeV (c) 180 GeV (d) 190 GeV from electron-positron collisions at the ILC at $\sqrt{s}$ = 500 GeV and 1000 fb$^{-1}$ integrated luminosity.}
\label{figure:fig3}
\end{figure}

\subsection{Monte Carlo simulation of the background}

The background samples were also produced using WHIZARD-3.4.1 with O’MEGA for matrix element calculation and VAMP for phase space integration. The electromagnetic and hadronic showers were simulated using PYTHIA6 and the detector simulation was done using DELPHES for the International Large Detector (ILD) at the ILC.

The first background process is the charged dilepton pair production: Drell-Yan (DY) process ($e^+ e^- \rightarrow e^+ e^-$).  
The second source of background is the diboson pair production $WW$ and $ZZ$ where both are decayed leptonically. Also the top anti-top production $t\Bar{t}$ where the top decays to $W$ boson and $b$ quark and the $W$ is decayed leptonically. 


\begin{table} [t]
\centering
\begin {tabular} {|l|l|l|c|l|}
\hline
Process \hspace{1pt} & Generator  & final state & $\sigma \times~\text{BR} ~(\text{fb})$ & Order \\
\hline
\hline
$\text{DY}  $  & WHIZARD & $e^{-}e^{+}$  & 1.685$\times 10^{5}$ & LO\hspace{6pt}\\
\hline
$\text{t}\bar{\text{t}}$ & WHIZARD  &$e^{+} e^{-}+2b+2\nu_e$ &9.46$\times 10^{0}$ & LO \\ 
\hline
$WW$  & WHIZARD &$e^{+} e^{-}+2\nu_e$ & 1.8$\times 10^{2}$ & LO \\
\hline
$ZZ$  & WHIZARD  & $2e^{+} + 2e^{-}$ &3.72$\times 10^{0}$& LO \\
\hline
$ZZ$  & WHIZARD &$e^{+} e^{-}+2\nu_e$ & 4.72$\times 10^{-1}$& LO \\
\hline
\end {tabular}
\vspace{3pt}
\caption{The SM background samples for the ILC at $\sqrt{s} = 500$ GeV. The corresponding cross-section times branching ratio for each process is also provided, along with the order of calculations, the final state, and the generator used.
}
\label{table:tab3}
\end{table}

\section{Event selection}
\label{section:MCandDat}

The pre-selection strategy for the chosen events is based on the presence of two oppositely charged leptons and a significant missing transverse energy. 
Each lepton in the event must have transverse momentum ($p^{e}_T$) greater than 10 GeV, a pseudo-rapidity ($|\eta^{e}|$) less than 3 radians, and an isolation variable 
($\sum_ip_T^i /p_T^{e}$) less than 0.1. These pre-selection cuts are listed in table \ref{table:tab3}. 

Figure \ref{figure:pfmet_bef_ele} shows the missing transverse energy ($E_{T}^{miss}$) distributions for signal and several SM background events 
after the pre-selection phase listed in table \ref{table:tab3}.  
The background histograms are stacked with the Drell-Yan events in cyan, 
the sum of $WW$ and $ZZ$ (denoted by EWK) in yellow, 
and the $t\bar{t}$ events in red.
The signal for $\lambda_L$= 0.1 is overlaid 
in the solid black line for the vector-like lepton mass of 200 GeV 
and dark matter masses of  1, 100, 180 and 190 GeV. 
All events were generated from electron-positron collisions at the ILC 
with $\sqrt{s}$ = 500 GeV, 
and are normalized to their cross-sections at 1000 fb$^{-1}$ integrated luminosity.

\begin{table} [t]
\centering
\begin {tabular} {|c|c|c|}
\hline
      $\text{Pre-selection}$ & $\text{Final selection}$\\
\hline
\hline
        
                        $p_{T}^e$  > 10 GeV & $p_{T}^e$  > 10 (GeV)\\   
                         $ |\eta^e| $ < 3 (rad) & $ |\eta^e| $ < 3 (rad) \\
                         $\sum_ip_T^i /p_T^{e}$ < 0.1 & $\sum_ip_T^i /p_T^{e}$ < 0.1   \\

     &$|p_{T}^{e^{+}e^{-}} -E_{T}^{miss}|/p_{T}^{e^{+}e^{-}} < 0.2$  \\ 
                            &$\Delta R_{ e^+ e^-} < 3.2$   \\
                             & Leading electron $|\eta^e_1|$ < 0.7   \\
                    
\hline
\end {tabular}
\vspace{3pt}
\caption{Summary of pre-selection and final selection cuts.  
}
\label{table:tab3}
\end{table}
\begin{figure*}
    \centering
    \begin{subfigure}{0.65\textwidth}
        \includegraphics[width=\textwidth]{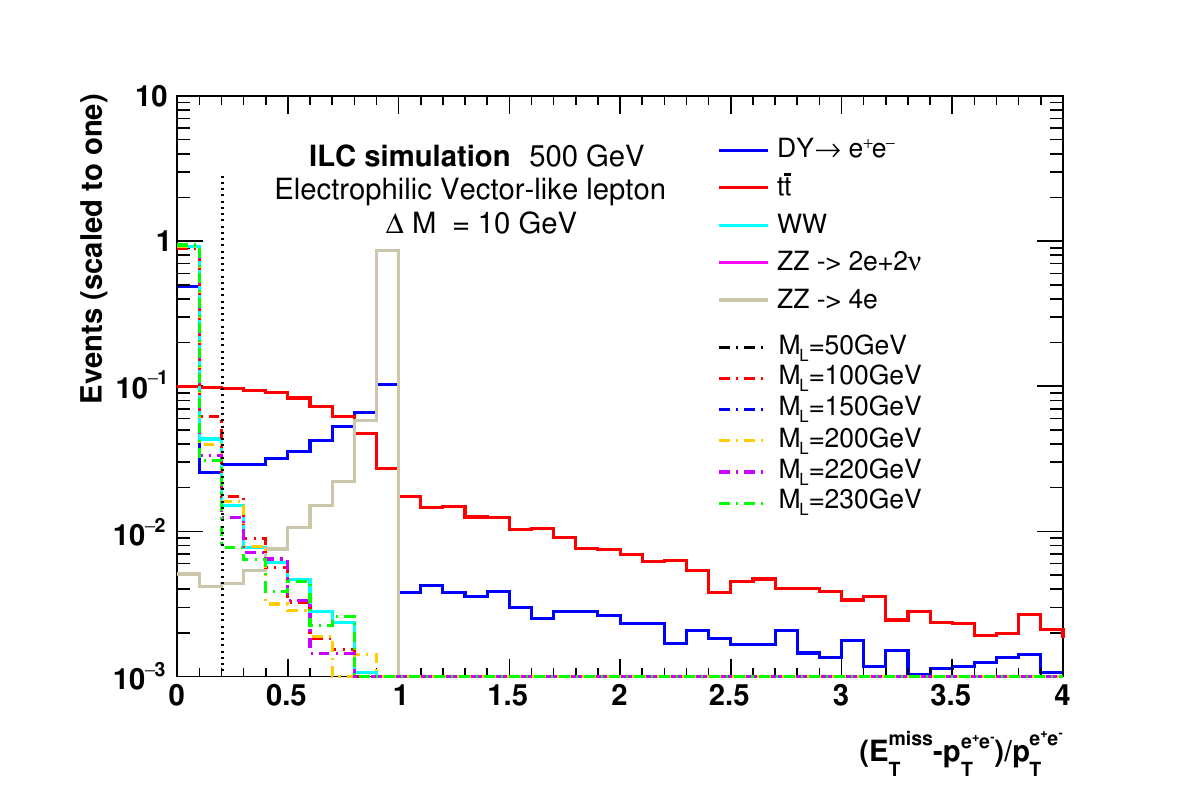}
        \caption{Relative difference between $P_{T}$ of the dielectron and $E_{T}^{miss}$}
        \label{fig:diff_plot_deltapt}
    \end{subfigure}
    \begin{subfigure}{0.65\textwidth}
        \includegraphics[width=\textwidth]{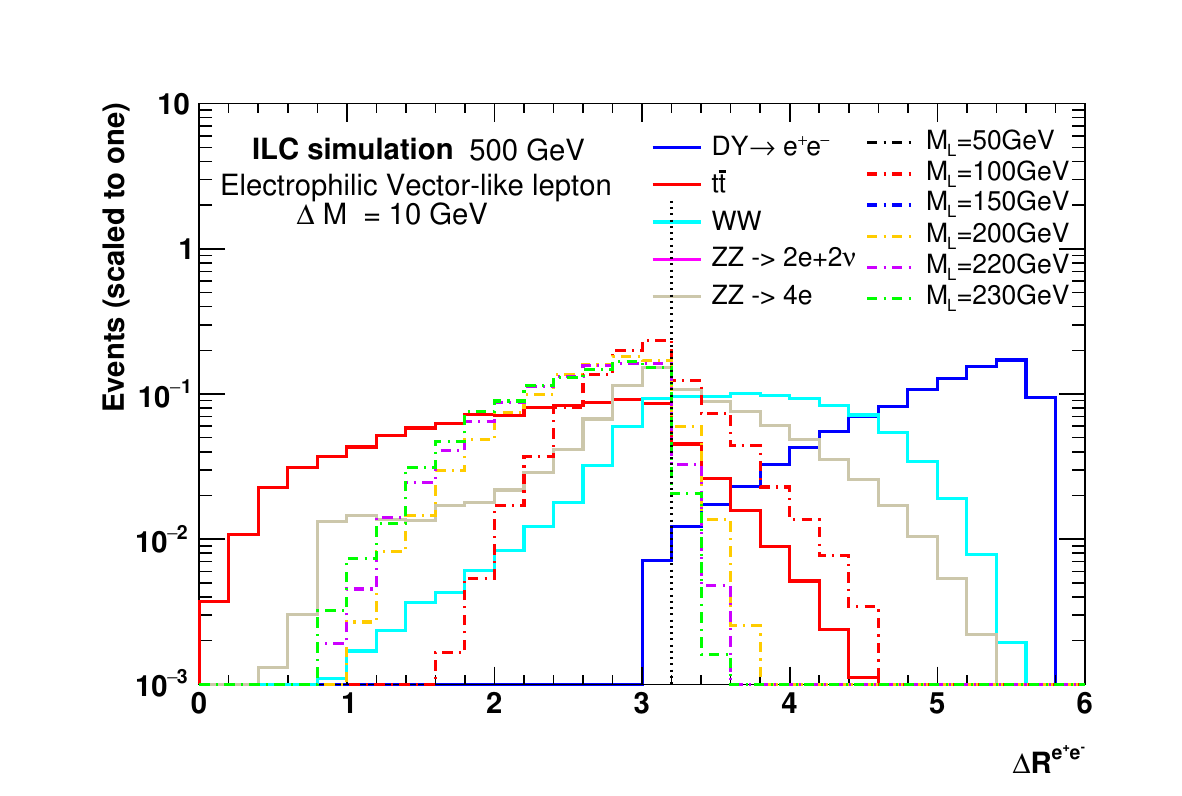}
        \caption{The distance in the ($\eta,\phi$) plane between the two electrons .}
        \label{fig:diff_plot_deltaR}
    \end{subfigure}
    \begin{subfigure}{0.65\textwidth}
        \includegraphics[width=\textwidth]{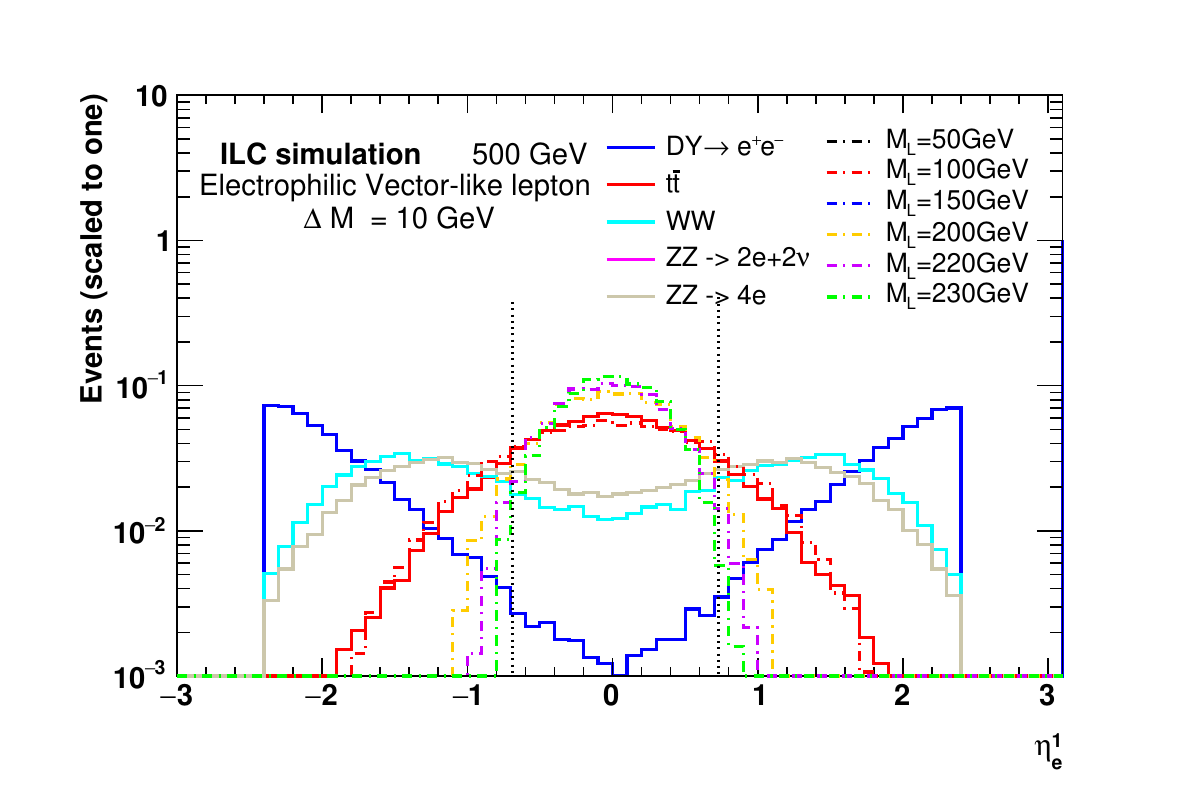}
        \caption{Pseudo-rapidity of the leading electron}
        \label{fig:diff_plot_eta}
    \end{subfigure}

    \caption{Difference plots between the signal 
    at various vector-like lepton masses for mass splitting 
    of $\Delta M = 10$~GeV and background samples are scaled to one. }
    \label{fig:fig4}
\end{figure*}

The following cuts, summarized in Table~\ref{table:tab3}, 
have been implemented to refine the analysis: a reduction in the difference between the transverse energy of the dilepton and the missing transverse energy to less than 0.2 times the transverse energy of the dilepton itself; a limitation on the distance between the two leptons in the ($\eta,\phi$) plane to less than 3.2 units; and 
a cut on the pseudorapidity of the leading lepton to be less than 0.7.

In figure \ref{fig:fig4}, 
we show the simulated background samples and the 
signals for each kinematic variable used for the final selection. 
The signal represents the dashed lines, 
while the solid lines represent the background. 
All the histograms are scaled to one. 
The plots cover vector-like lepton masses from $50-230$ GeV.

\begin{figure}[t]
\centering
\begin{subfigure}[b]{0.45\textwidth}
\centering
\includegraphics[width=\textwidth]{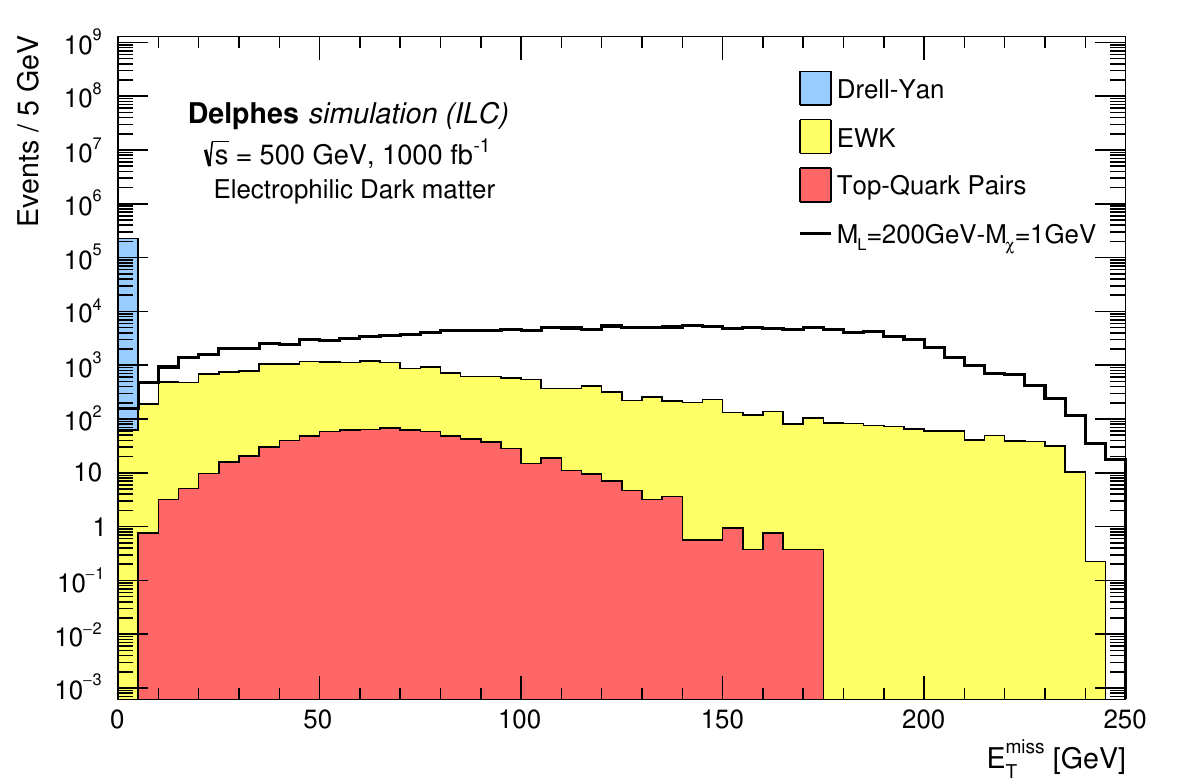}
\caption{}
\label{figure:pfmet_bef_ele1}
\end{subfigure}
\begin{subfigure}[b]{0.45\textwidth}
\centering
\includegraphics[width=\textwidth]{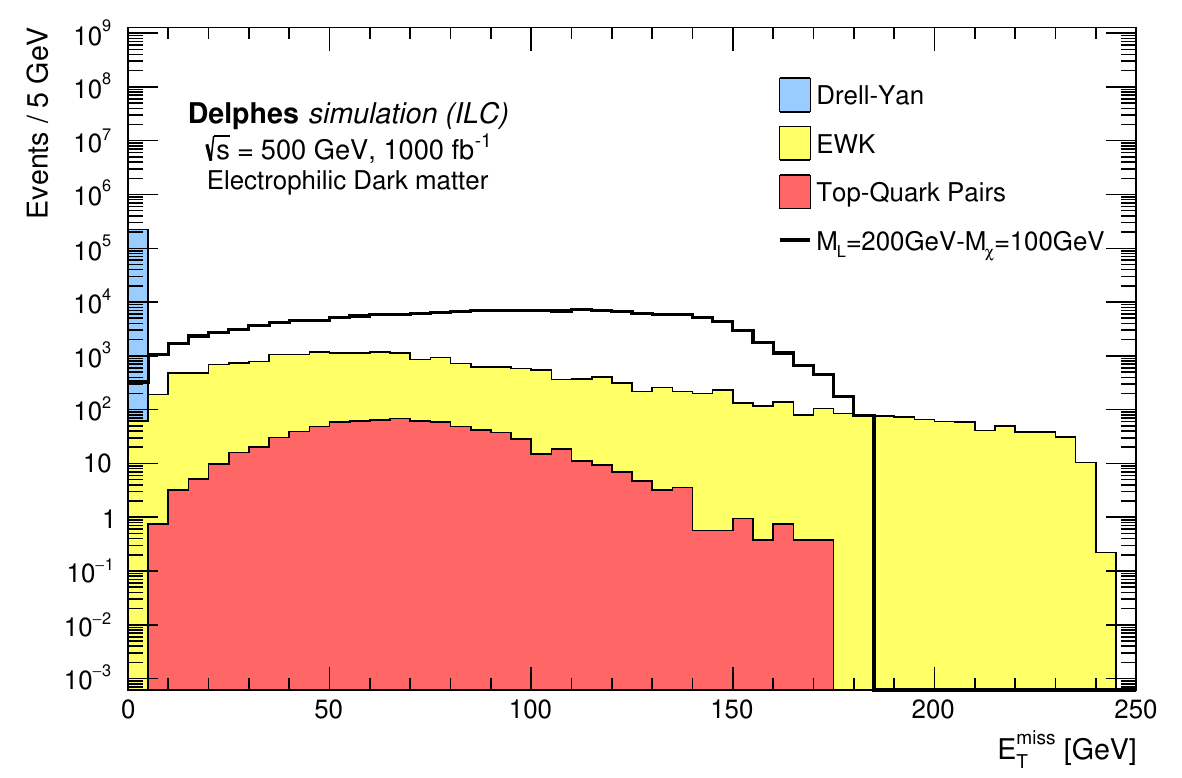}
\caption{}
\label{figure:pfmet_bef_ele2}
\end{subfigure}
\begin{subfigure}[b]{0.45\textwidth}
\centering
\includegraphics[width=\textwidth]{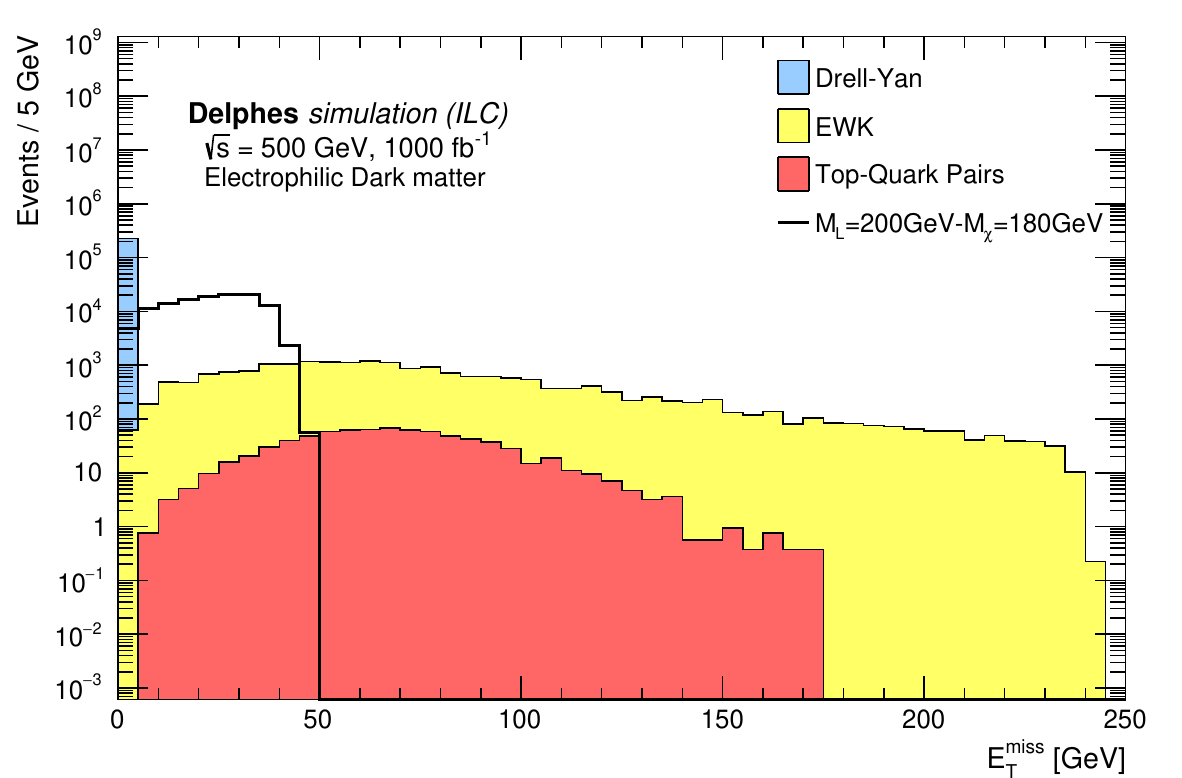}
\caption{}
\label{figure:pfmet_bef_ele3}
\end{subfigure}
\begin{subfigure}[b]{0.45\textwidth}
\centering
\includegraphics[width=\textwidth]{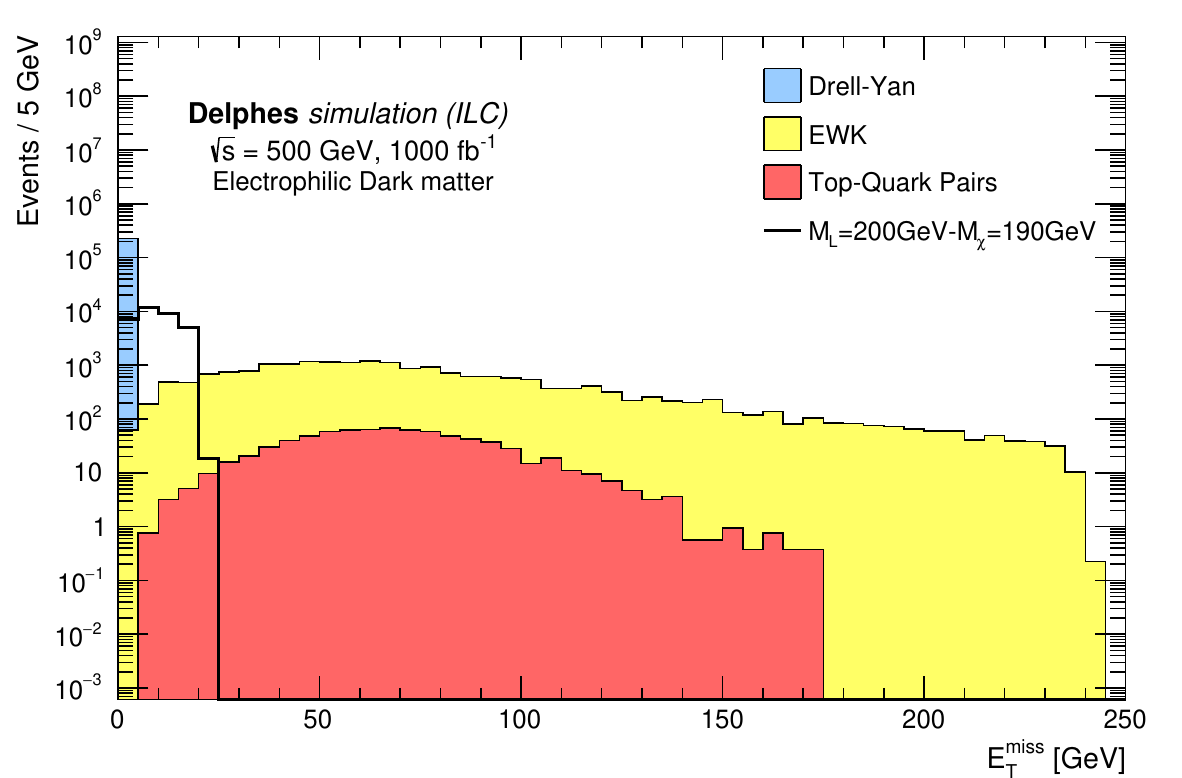}
\caption{}
\label{figure:pfmet_bef_ele4}
\end{subfigure}
\caption{The missing transverse energy spectrum, for events passing the final-selection cuts listed in table \ref{table:tab3}, of the estimated SM backgrounds together with the LPDM for electro-philic vector-like lepton of mass  200 GeV and scalar dark matter mass of (a) 1 GeV (b) 100 GeV (c) 180 GeV (d) 190 GeV from electron-positron collisions at the ILC at $\sqrt{s}$ = 500 GeV and 1000 fb$^{-1}$ integrated luminosity.}
\label{figure:fig5}
\end{figure}


\section{Results}
\label{section:results}

The results of the study were concluded by using a shape-based analysis 
where the $E_{T}^{miss}$ is used as the discriminating variable. 
The distributions of missing transverse energy are showcased 
in figure \ref{figure:fig5} for both SM backgrounds 
and signal concerning vector-like lepton mass $M_{\VLL} = 200$ GeV 
and dark matter mass $M_{\chi}$= 1, 100, 180, 190 GeV. 
The distributions are presented for the electro-philic case 
with $\lambda_L = 0.1$ generated from electron-positron collisions at the ILC. The collisions are conducted at a $\sqrt{s}$ of 500 GeV with an integrated luminosity of 1000 fb$^{-1}$, 
and the events are selected by the final cuts that are summarized in table \ref{table:tab3}.
%
%
%
%
%
%
%
%
We calculated the 95\% C.L. exclusion limits on the cross-section, 
using the profiled likelihood ratio test statistic 
based on the modified frequentist method \cite{stats1,stats2}. We used the asymptotic approximation \cite{stats3} and separately computed limits for each mass point of the $L$. 

\begin{figure*}
    \centering
    \begin{subfigure}[b]{0.65\textwidth}
    \includegraphics[width=\textwidth]{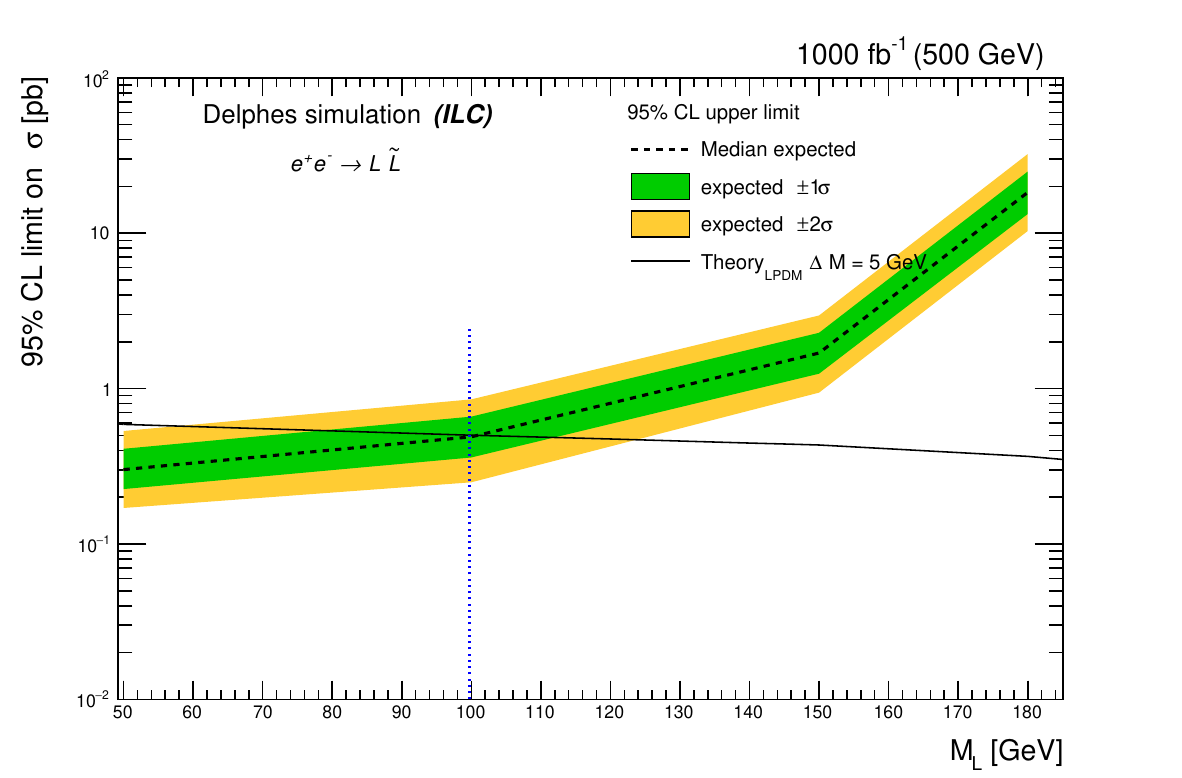}       
        \caption{}
        \label{fig:subfig1}
    \end{subfigure}
    \hfill
    \begin{subfigure}[b]{0.65\textwidth}
    \includegraphics[width=\textwidth]{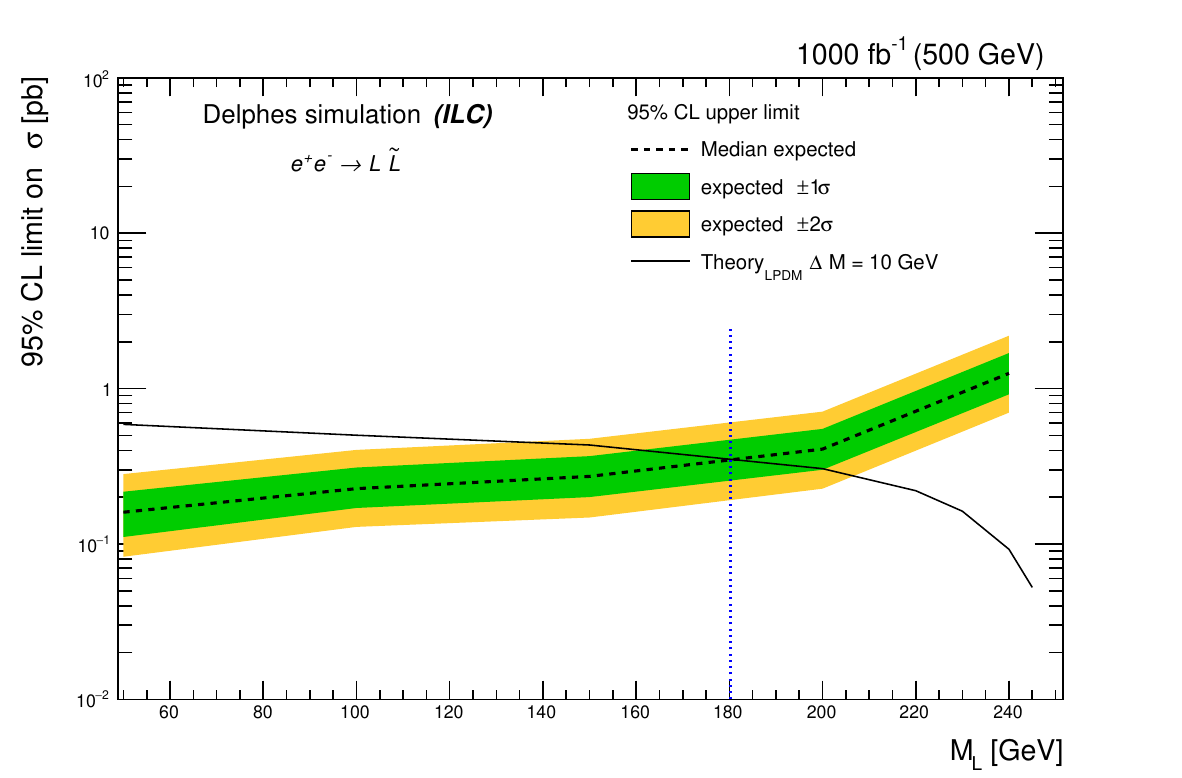} 
        \caption{}
        \label{fig:subfig2}
    \end{subfigure}
    \hfill
    \begin{subfigure}[b]{0.65\textwidth}
    \includegraphics[width=\textwidth]{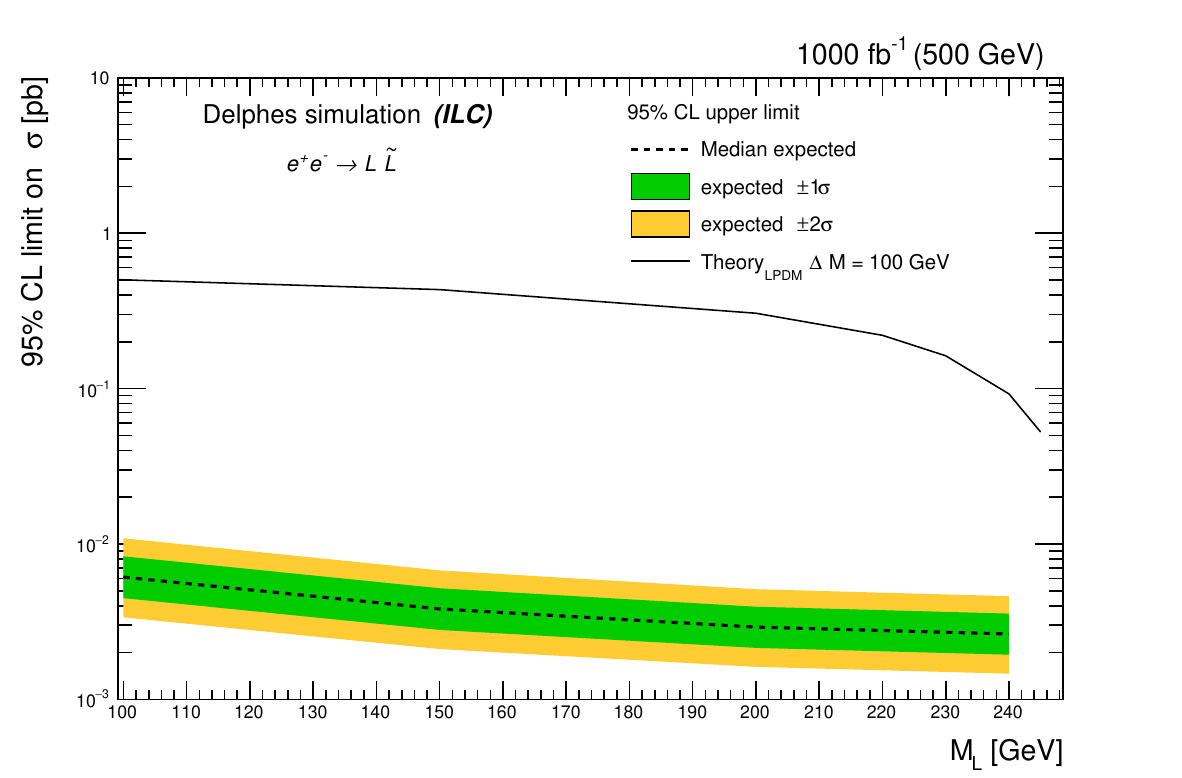} 
        \caption{}
        \label{fig:subfig3}
    \end{subfigure}
    \caption{
    95\% C.L. limit on the expected production cross-section of the vector-like leptons as a function of the vector-like lepton mass. 
    The solid line represents the cross-section predicted from the electro-philic vector-like lepton theory with a yukawa coupling of 0.1 and (a) $\Delta M$=5 GeV (b) $\Delta M$= 10 GeV (c) $\Delta M$= 100 GeV.  }
    \label{figure:fig6}
\end{figure*}

Figure \ref{figure:fig6} shows the 95\% confidence limit on the cross-section of the vector-like lepton simplified model with leptonic decay of $\VLL$. 
It presents the results for three scenarios: wide mass splitting $\Delta M$ = 100 GeV  (\ref{fig:subfig3}), mass splitting of $\Delta M$ = 10 GeV (\ref{fig:subfig2}), and narrow splitting $\Delta M$ = 5 GeV (\ref{fig:subfig1})   with $\lambda_L=0.1$ for all the cases. 
The lepton portal dark matter model predicts the cross-section 
as a black solid line. 
The statistical test provides the expected limit as a dashed line, 
while the green and yellow bands represent the $\pm1$ and $\pm2$ sigma bands.
The limit plots indicate clear exclusion limits on the mass of the vector-like leptons ($\VLL$) under various scenarios of mass splitting between the $\VLL$ and $\chi$ states. For a narrow mass splitting of $\Delta M = 5$ GeV, any $\VLL$ mass below 100 GeV is ruled out. When the mass splitting increases to $\Delta M = 10$ GeV, the exclusion limit on the $\VLL$ mass rises to below 180 GeV. In cases with significant mass splitting, the entire range of possible masses for the $\VLL$ is excluded. 
As noted in Ref.~\cite{lpdm}, a similar search strategy employed at the LHC cannot effectively probe the mass-degenerate region where $\Delta M$ is less than or equal to 80 GeV. Therefore, the International Linear Collider (ILC) is poised to play a crucial role in the search for vector-like leptons that are nearly degenerate with dark matter.




\section{Summary}
\label{section:summary}

The International Linear Collider (ILC) is a proposed lepton collider that would collide beams of electrons and positrons. The ILC will be capable of scanning the center of mass energies from 250 to 500 GeV, possibly increasing this to 1 TeV. The main aim of the ILC is to discover physics beyond the Standard Model.
The ILC is unique in that it can use polarized beams. This option could lead to a higher signal-to-background ratio, which would enhance the strength of the new physics signal.

In our analysis, we investigated the possibility of producing the lepton portal dark matter.   
We focused on the electron-philic scalar dark matter scenario, where the dark matter is created from the decay of extra vector-like leptons, that are produced in pairs from electron-positron collisions at the ILC with a center-of-mass energy of 500 GeV. 
These vector-like leptons can decay into a scalar dark matter and an electron. 
We have simulated signal and background samples produced from electron-positron collisions at the center of mass energy of 500 GeV and 1000 fb$^{-1}$ integrated luminosity corresponds to ILC Run I. 
The degree of polarization for the electron beam was set to 0.8 and for the positron beam to -0.3.

This analysis, based on the events that made it through the final selection, allowed us to set upper limits on the production cross-section. Additionally, we established an exclusion limit for the mass of the doublet vector-like lepton $L$. This was done under various scenarios for the mass splitting between the vector-like lepton and the dark matter, specifically with $\Delta M$ values of 5 GeV, 10 GeV, and 100 GeV, while assuming a Yukawa coupling of $\lambda_L$ = 0.1.
We presented the conserved case with a small Yukawa coupling constant, noting that sensitivities will be heightened for larger Yukawa couplings due to the inclusion of the additional T-channel process, as illustrated in figure \ref{figure:fig2}. In scenarios with a wide mass splitting of $\Delta M = 100$ GeV, we successfully excluded the entire range of $M_L$ under consideration. For a mass splitting of $\Delta M = 10$ GeV, we were able to rule out vector-like lepton masses up to 180 GeV. Additionally, in the case of a narrow mass splitting of $\Delta M = 5$ GeV, we excluded vector-like lepton masses up to 100 GeV. This study demonstrates the potential to investigate situations where vector-like leptons are nearly degenerate with dark matter, specifically when $\Delta M < 100$ GeV—something the LHC could not access.

The results for the muon-philic scenario would align closely with those found in this study, assuming the reconstruction efficiencies and resolutions for both electrons and muons are comparable. However, the sensitivities for the tau-philic case are expected to be weaker due to the complexities associated with tau signals in the detector. Detailed studies on these specific cases are, however, beyond the scope of this paper.

\begin{acknowledgments}
Y. Mahmoud wishes to acknowledge the support of the Center for Theoretical Physics (CTP) at the British University in Egypt (BUE) for the financial support to this work. 
In addition, this paper is based on works supported by the Science, Technology, and Innovation Funding Authority (STIF) under grant number 48289. 
The work of J. Kawamura was supported by IBS under the project code, IBS-R018-D1.
\end{acknowledgments}

\nocite{*}


\end{document}